\documentclass[journal=nalefd,manuscript=communication]{achemso}
\usepackage[version=3]{mhchem}
\usepackage{colordvi}

\author{Yafei Ren}
\affiliation{ICQD, Hefei National Laboratory for Physical Sciences at Microscale, and Synergetic Innovation Center of Quantum Information and Quantum Physics, University of Science and Technology of China, Hefei, Anhui 230026, China.}
\alsoaffiliation{CAS Key Laboratory of Strongly-Coupled Quantum Matter Physics, and Department of Physics, University of Science and Technology of China, Hefei, Anhui 230026, China.}
\author{Ke Wang}
\affiliation{ICQD, Hefei National Laboratory for Physical Sciences at Microscale, and Synergetic Innovation Center of Quantum Information and Quantum Physics, University of Science and Technology of China, Hefei, Anhui 230026, China.}
\alsoaffiliation{CAS Key Laboratory of Strongly-Coupled Quantum Matter Physics, and Department of Physics, University of Science and Technology of China, Hefei, Anhui 230026, China.}
\author{Xinzhou Deng}
\affiliation{ICQD, Hefei National Laboratory for Physical Sciences at Microscale, and Synergetic Innovation Center of Quantum Information and Quantum Physics, University of Science and Technology of China, Hefei, Anhui 230026, China.}
\alsoaffiliation{CAS Key Laboratory of Strongly-Coupled Quantum Matter Physics, and Department of Physics, University of Science and Technology of China, Hefei, Anhui 230026, China.}
\author{Shengyuan A. Yang}
\affiliation{Research Laboratory for Quantum Materials, Singapore University of Technology and Design, Singapore 487372, Singapore}
\author{Jeil Jung}
\affiliation{Department of Physics, University of Seoul, Seoul 02504, Korea}
\email{jeiljung@uos.ac.kr}
\author{Zhenhua Qiao}
\affiliation{ICQD, Hefei National Laboratory for Physical Sciences at Microscale, and Synergetic Innovation Center of Quantum Information and Quantum Physics, University of Science and Technology of China, Hefei, Anhui 230026, China.}
\alsoaffiliation{CAS Key Laboratory of Strongly-Coupled Quantum Matter Physics, and Department of Physics, University of Science and Technology of China, Hefei, Anhui 230026, China.}
\email{qiao@ustc.edu.cn}
\title{Gate tunable current partition in graphene based topological zero lines}

\begin{document}
\begin{abstract}
  We demonstrate new mechanisms for gate tunable current partition at topological zero-line intersections in a graphene-based current splitter. Based on numerical calculations of the non-equilibrium Green's functions and Landauer-B\"{u}ttiker formula, we show that the presence of a perpendicular magnetic field on the order of a few Teslas allows for carrier sign dependent current routing. 
  In the zero-field limit the control on current routing and partition can be achieved within a range of $10\%$-$90\%$ of the total incoming current by tuning the carrier density at tilted intersections, or by modifying the relative magnitude of the bulk band gaps via gate voltage. We discuss the implications of our findings in the design of topological zero-line networks where finite orbital magnetic moments are expected when the current partition is asymmetric.
\end{abstract}

\textbf{Introduction}.
Topological charge transport channels can be tailored at the interfaces of insulating systems with topologically nontrivial bulk gaps~\cite{QAHE1,qiaoti,TopologicalStates0,TopologicalStates1,TopologicalStates2,TopologicalStates3,kanemele, QAHE2,QAHE3,QAHE4,QAHE5,QAHE6,QAHE7,QAHEExp1,QAHEExp2,QAHEExp3,QAHEExp4}. Traditionally, the honeycomb lattices of graphene single and multilayers have provided a suitable platform that support a variety of quantum Hall phases including the anomalous Hall~\cite{QAHE1}, valley Hall~\cite{valleyhall}, spin Hall~\cite{kanemele,qiaoti} effects depending on their relative spin-valley flavor dependent gap signs~\cite{QAHE7,junghf}. At the domain wall interfaces between different bulk gap signs, in particular for the one-dimensional zero-line modes appearing at the interface between regions with opposite valley Hall conductivities~\cite{Martin,yao,semenoff,Jeil,eunah,yongkuk,BiXintao,changhee}, numerical studies have shown transport properties that are robust against backscattering, leading to practically ballistic transport through zero-line modes with mean free paths of the order of hundreds of microns in relatively clean samples~\cite{Highway,zhujun}. This remarkable robustness against backscattering would be preserved in the presence of turns and bends of the zero lines suggesting that essentially arbitrary trajectories would preserve excellent transport qualities. Recent years have witnessed experimental progresses towards the realization of these zero-line channels through careful gate alignment~\cite{zhujun,hujong} or in samples of bilayer graphene with stacking faults~\cite{julong}. While the field is yet in its infancy, it is expected that research of transport along topological zero lines will thrive in a near future.

A somewhat more complex but rich device setup consists of  two intersecting zero lines, along which the injected current can partition~\cite{Highway}. The outgoing current is in principle allowed to propagate only when the relative signs of the surrounding valley Hall gaps preserve the chirality of the incoming current. In a perpendicular intersection, the current injected through one of the leads is partitioned evenly into two outgoing channels, whereas in two intersecting zero lines with tilted angles, the current partition follows a rule analogous to the Snell's refraction laws of conventional optical beam splitters~\cite{partition}. Numerical studies have shown that these zero-line beam splitters consisting of nodes of intersecting zero lines are expected to preserve the current partition properties for rather strong disorder potentials that are comparable to the size of the band gap~\cite{Highway}, suggesting that these current partition behavior can be present in actual devices.

In this Letter, we present a new proposal for current routing in a zero-line beam splitter setup that is controllable through a single backgate, which takes advantage of a rather special carrier-dependent tunability of the current partition laws at zero-line intersections. We identify new control knobs that can tune the current partition in addition to the zero-lines intersection angle reported in an earlier work~\cite{partition}.
The system parameters that introduce changes in the current partition laws include the (a) Fermi energy, (b) different relative masses, and (c) perpendicular magnetic field, all of which can be more easily modified in a device than the intersection angles. In the following, we discuss briefly the system setup of intersecting zero lines based on graphene before we move on to investigate the influence in the current partition of these newly proposed system control parameters.

\begin{figure}
  \includegraphics[width=8 cm,angle=0]{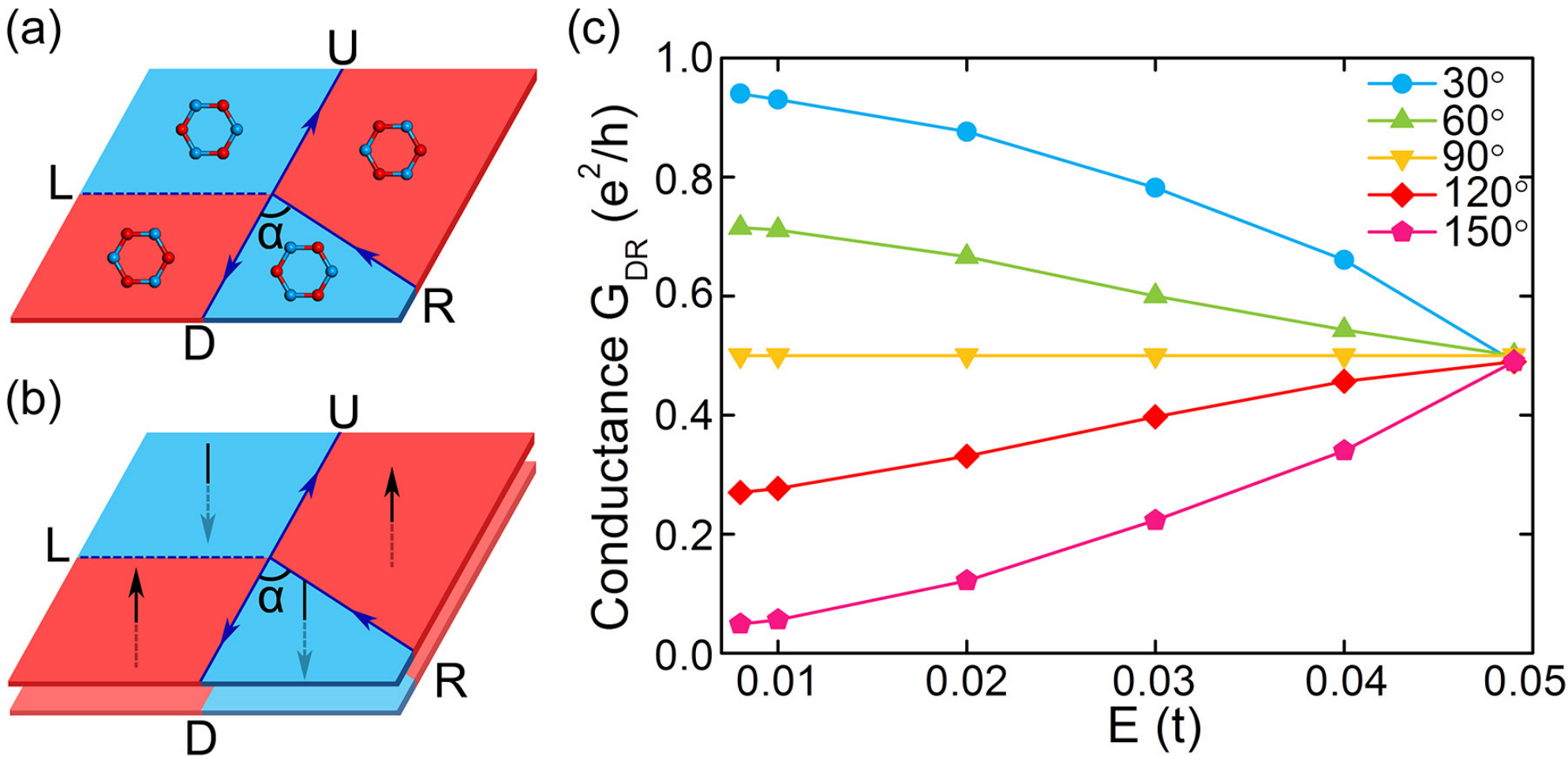}
  \caption{Schematic plots of four-terminal zero-line channel splitters as a function of incoming angle and carrier density. Current injected from lead R can only split into two streams propagating along leads U and D, while the propagation via lead L is completely forbidden due to chirality preservation rules. The lead R forms an angle $\alpha$ with respect to lead D. (a) Alternating sublattice potentials are considered at the four connecting quadrants in a monolayer honeycomb lattice; (b) Alternating perpendicular electric fields are imposed at the four connecting quadrants in a bilayer graphene. (c) Current partition dependence on the Fermi level for different angles of incidence $\alpha$. Five representative angles of incidence $\alpha=30^\circ,60^\circ,90^\circ,120^\circ$ and $150^\circ$ are considered. When the Fermi level approaches the bulk band edge, the current tends to partition equally towards the outgoing leads.}
  \label{Fig1}
\end{figure}

\textbf{Intersecting zero-lines}.
Figure~\ref{Fig1} shows the schematic plots of four-terminal zero-line electron beam splitters formed in (a) monolayer honeycomb lattice with staggered sublattice potentials denoted by red or blue and (b) bilayer graphene experiencing different vertical electric fields indicated by the out-of-plane arrows that give rise to mass terms and opens up band gaps at K and K$'$ points. At the interfaces between regions of opposite mass signs, zero lines are formed, along which the current flows as labelled by blue arrows. We consider the following geometry in our numerical calculations: the up (U) and down (D) zero lines are fixed along the vertical direction, the left (L) zero line is fixed along the horizontal direction and perpendicular to UD, while the right (R) segment of the zero line remains free to form an angle $\alpha$ with respect to the D zero line. In the following, we present our analysis of the transport properties on the single layer graphene with mass terms due to staggered sublattice potentials. We expect that the same conclusions will hold for other chiral two dimensional gas systems like Bernal stacked bilayers as verified in earlier work~\cite{Highway}. The corresponding tight-binding Hamiltonian for the system we study is written as $H =-t\sum_{\langle ij \rangle} ~ c_i^{\dag} c_j + \lambda\Delta (\sum^A_{i} c_{i}^{\dag} c_{i} - \sum^{B}_{i} c_{i}^{\dag} c_{i})$, where $c^{\dag}_i$~($c_{i}$) is a creation~(annihilation) operator for an electron at site $i$, $t=2.6~$eV is the magnitude of the nearest neighbor hopping amplitude, $\Delta$ measures the staggered potential strength and $\lambda$=``$+--+$'' specifies the sign of the gap in each one of the four quadrants, which in turn determines the sign of the valley Hall effect~\cite{valleyhall}. Throughout this paper, we use the potential size of $\Delta/t=0.05$ unless otherwise specified.

\textbf{Fermi-level tuning}.
At the charge neutrality point, the current partition can be explicitly determined by a function that depends only on their intersection angles~\cite{partition}. Here we show that away from the charge neutrality point the current partition laws are modified when the zero-line segments make an angle different from the simple case of 90$^{\circ}$. How the partition laws can be modified as a function of Fermi energy is shown in Fig.~\ref{Fig1}(c) where $G_{\rm DR}$ represents the conductance for currents injected through the R lead flowing towards the D lead. Bearing in mind that $G_{\rm DR} + G_{\rm UR} = e^2/h$ for a spinless system, we note that near charge neutrality a greater amount of current partitions towards the lead that has a sharper turn angle. When the Fermi level is shifted away from the neutrality point towards the bulk gap edges $E_{\rm F} \simeq \pm \Delta$, we find that the conductances gradually reach equipartition $G_{\rm DR} = G_{\rm UR} = e^2/2h$. 

\begin{figure}
  \includegraphics[width=8 cm,angle=0]{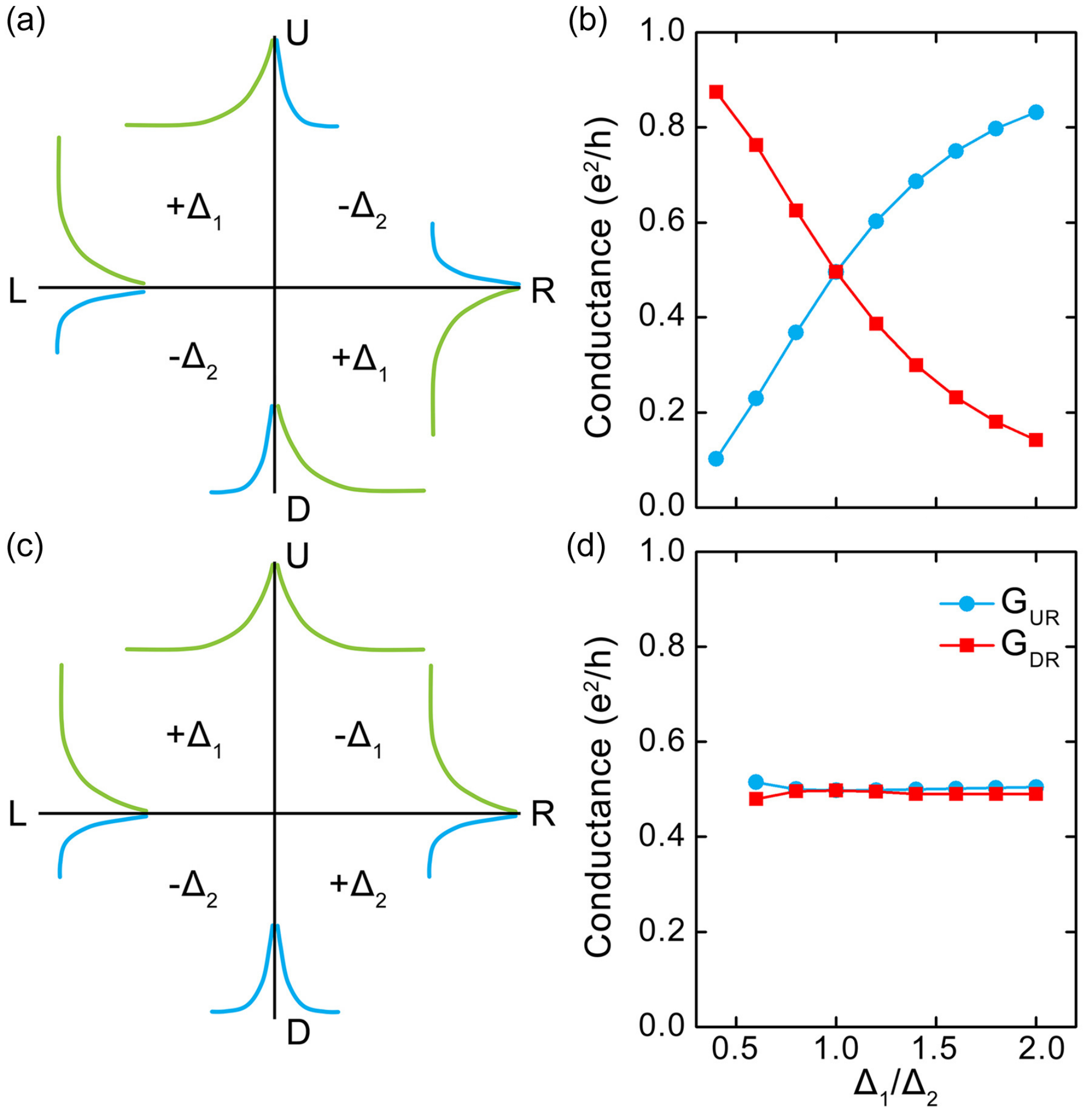}
  \caption{The dependence of current partition on relative masses between neighboring quadrants in a four terminal zero-line splitter. Here, the system structure is composed of two crossing zero lines with $\alpha=90^\circ$. (a) and (c): The alternating masses are respectively ``$+\Delta_1,-\Delta_2,-\Delta_2,+\Delta_1$" and ``$+\Delta_1,-\Delta_1,-\Delta_2,+\Delta_2$". We use green and blue curves to schematically show the decay of wavefunctions away from the zero lines in the regions with $\Delta_1$ and $\Delta_2$, respectively, by assuming $\Delta_1<\Delta_2$. (b) and (d): The corresponding conductances from right to up and down leads $\rm G_{UR}$ and $\rm G_{DR}$ as a function of the relative strength $\Delta_1/\Delta_2$ under two different potential profiles as shown in (a) and (c).}
  \label{Fig2}
\end{figure}

\textbf{Relative mass variation}.
A different system parameter that can change the partition of the currents is the relative magnitude of the mass terms that defines the zero line. 
As pointed out in Ref.\citenum{partition}, the wavefunctions of the zero-line modes are widely spread across the zero line, and the overlap between the incoming and outgoing wavefunctions can be counted as the main factor that determines the current splitting. 
The spatial decay of the zero-line-mode wave functions into the bulk follows an exponential law $\propto e^{- x/x_0}$ in the direction $x$ perpendicular to the zero line where $x_0 \sim 1/\Delta$ dictating the length scale of the tail is inversely proportional to the size of the band gap $\Delta$ in the bulk~\cite{partition}. One natural conclusion is that the relative magnitude of the band gaps surrounding the incoming and outgoing leads should allow to change the ratio of the partitioned currents into U and D leads. For sake of clarity, we focus our discussions for a setup with the angle between R and D being $\alpha=90^\circ$ where all contiguous leads are perpendicular to each other.
We adopt two different mass sizes in the four quadrants that we label as $\Delta_1$ and $\Delta_2$. In Figs.~\ref{Fig2}(a) and  \ref{Fig2}(c), we show two different configurations for the mass sizes, i.e., $``+\Delta_1,-\Delta_2,-\Delta_2,+\Delta_1"$  and $``+\Delta_1, -\Delta_1, -\Delta_2, +\Delta_2"$, and the corresponding current partitions are represented as a function of the relative mass strength $\Delta_1/\Delta_2$ as shown in Figs.~\ref{Fig2}(b) and \ref{Fig2}(d), respectively. One can observe that, for the former case, the conductances $\rm G_{UR}$ and $\rm G_{DR}$ are highly sensitive to the relative gap sizes $\Delta_1/\Delta_2$, e.g., when $\Delta_1>\Delta_2$ ($\Delta_1<\Delta_2$), much more current goes through the up (down) zero line. We discuss here the specific example of $\Delta_1 < \Delta_2$. As mentioned earlier, the wavefunction decays across the zero line into the bulk exponentially with a tail length depending on the inverse of the mass $\Delta^{-1}$. As shown schematically in Fig.~\ref{Fig2}(a), for the $``+\Delta_1,-\Delta_2,-\Delta_2,+\Delta_1"$ setup, the wavefunction decays faster in the larger mass region (blue lines) making the current incoming from the right lead spread more widely in the smaller mass $\Delta_1$ region. At the bifurcation point, there is more overlap between the wave functions incoming from the right and then turning to the down lead than for the turn to the up lead.

The situation is different for the second example $``+\Delta_1,-\Delta_1,-\Delta_2,+\Delta_2"$ where the partition of the current incoming from R remains equal for both U and D leads as indicated in the lower panel of Fig.~\ref{Fig2}(b). This can be understood if we keep in mind that the wavefunction from R lead overlap with an equal ratio for both U and D leads distributing equally with a $50\%$-$50\%$ partition despite that the spatial distribution of the wave functions are different on each side of the zero line. By further making use of the relations $\rm G_{RU}=G_{LD}$, $\rm G_{LU} = G_{RD}$ and $\rm G_{ij}=G_{ji}$~\cite{partition}, it is guaranteed that the current from any zero line must be equally partitioned. These observations indicate that the overlap ratio between the incoming and outgoing wave functions is of major relevance for the determination of the current partition laws.

\begin{figure}
  \includegraphics[width=16 cm,angle=0]{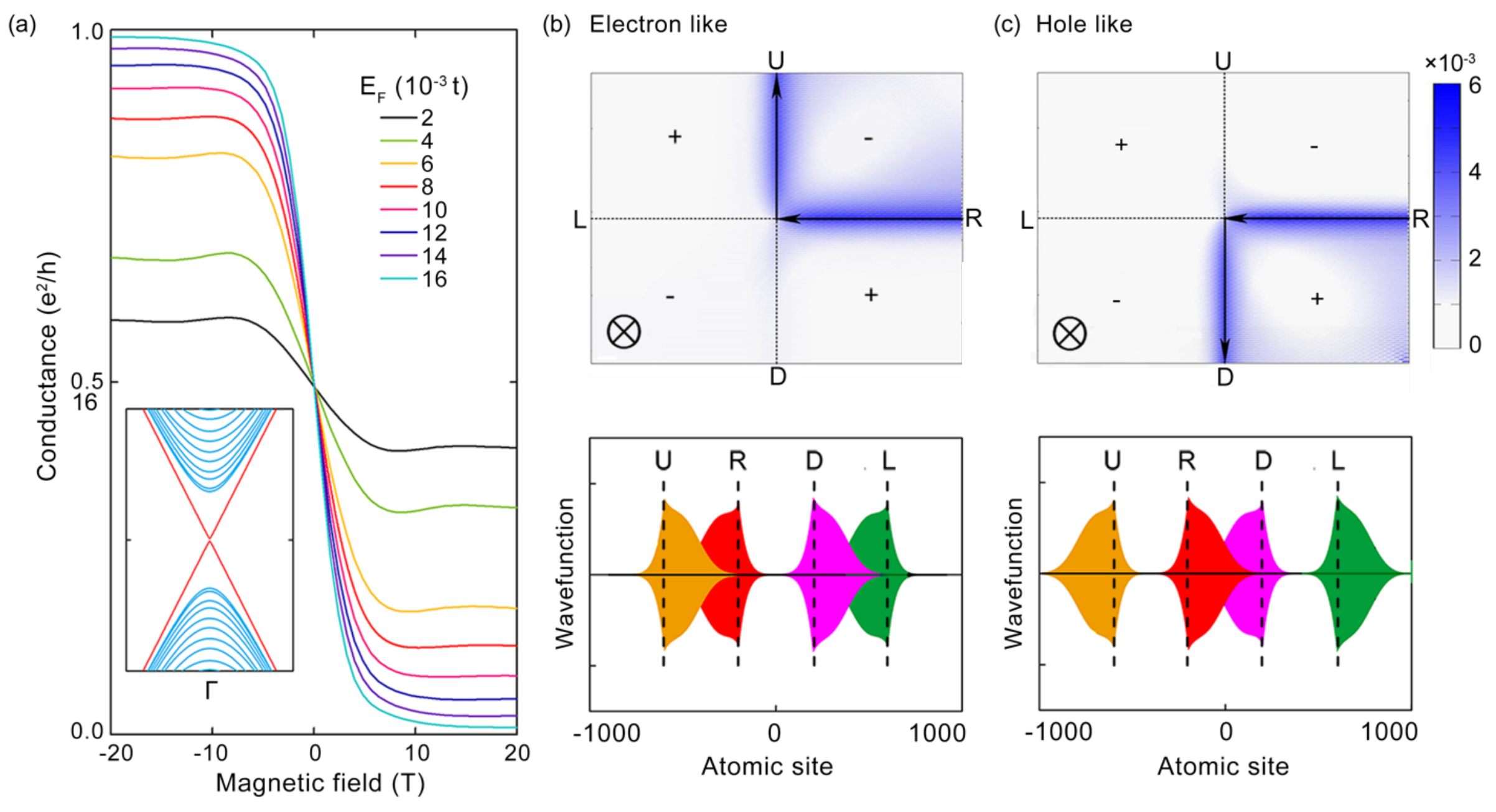}
  \caption{Influence of an external magnetic field on the current partition of a $\alpha=90^\circ$ zero-line node for different Fermi levels. (a) In the presence of a perpendicular magnetic field, the current partition asymmetry grows larger up to almost complete saturation when we modify the Fermi energy. Different curves correspond to different Fermi levels as indicated in the legend. The panels (b) and (c) show the local density of states of the injected currents normalized to one for electron-like and hole-like carrier densities in the presence of a perpendicular magnetic field of 8 Teslas. The opposite partition behaviors depending on carrier sign indicates the possibility of effectively routing the injected current by using electrical means, e.g. the backgate voltage. The band gap here is set to be $0.04~t$.}
  \label{Fig3}
\end{figure}

\textbf{Influence of a perpendicular magnetic field}.
A third control knob that influences the current partition for this system consists in an external perpendicular magnetic field. Once again we focus on the setup with $\alpha=90^\circ$ and set the masses at all quadrants to be equal, i.e., $``+\Delta,-\Delta, -\Delta,+\Delta"$ for sake of simplicity and to emphasize the influence of magnetic field on the current partition. The perpendicular magnetic field is applied evenly in the whole system, including the central scattering regime and all the four terminals. The vector potential is chosen to be ${\bf A} = (-By,~0,~0)$ for the horizontal leads and the central regime, so that the system has the translational symmetry along the $x$ direction, whereas for the U and D vertical leads we set the vector potential to be ${\bf A} = (0,~Bx,~0)$ when recursively calculating the the self-energies. A proper gauge transformation is included so that the self-energies from U, D leads are compatible with those in the
central regime as well as the L, R leads~\cite{qiao-magneticfield}. In Fig.~\ref{Fig3}, we show the conductance $\rm G_{UR}$ as a function of magnetic field $B$ for different Fermi energies $E_F$, where one can find that, when $B=0$, the current is equipartitioned to both U and D leads for every value of $E_F$ but this symmetry is broken as soon as a small magnetic field is applied. The asymmetry in the partition increases when we carrier dope the system. In particular, when the Fermi level approaches the bulk band gap edges, the current tends to saturate to either the U or D lead depending on the direction of the magnetic field and carrier doping, see Fig.~\ref{Fig3}.

Insight about this partition asymmetry can be gained if we consider that the center of the soliton wave packet on U or D is shifted to either right or left due to the Lorentz force acting on the charge density. This lateral displacement of the wave packet in the presence of a magnetic field breaks the even current partition symmetry towards the U and D leads, see the Supporting Information for more details. Moreover, as we change the Fermi energy away from the middle of the bulk band gap, the shift of the soliton wave functions is enhanced, which further increases the asymmetry of the wave function overlap and thus the current partition until eventual saturation. It is noteworthy that the magnetic-field assisted partition would act in opposite senses for the electron-like or hole-like cases as shown in Figs.~\ref{Fig3}(b) and \ref{Fig3}(c) where the Fermi energy is above (below) the middle of bulk band gap for the former (latter) case. We can find that the current partition
in the presence of an out-of-plane magnetic field, the electron (hole)-like doped system, more current is partitioned into U (D) lead in consistent with Lorentz force. In the lower panels of Figs.~\ref{Fig3}(b) and \ref{Fig3}(c), we show the wavefunction distributions of electron-like and hole-like states under magnetic field where one can clearly find that, for the electron-like case, the magnetic field moves the wavefunctions of R and U leads towards each other and thus the current tends to partition into U lead. The partition is opposite for the hole-like case. This carrier-type dependent current partition in the presence of a uniform magnetic field makes it possible to exercise control of current partition through external gating, which is preferable in practical device applications since the operation with electric signals has a faster switching speed than reversing the magnetic fields due to limitations from the circuit's inductance. Therefore, a device setup under a constant magnetic field and variable carrier densities through an external gate would in principle allow efficient current partition control by switching the propagation direction of the majority carriers.

\begin{figure}
  \includegraphics[width=16 cm, angle=0]{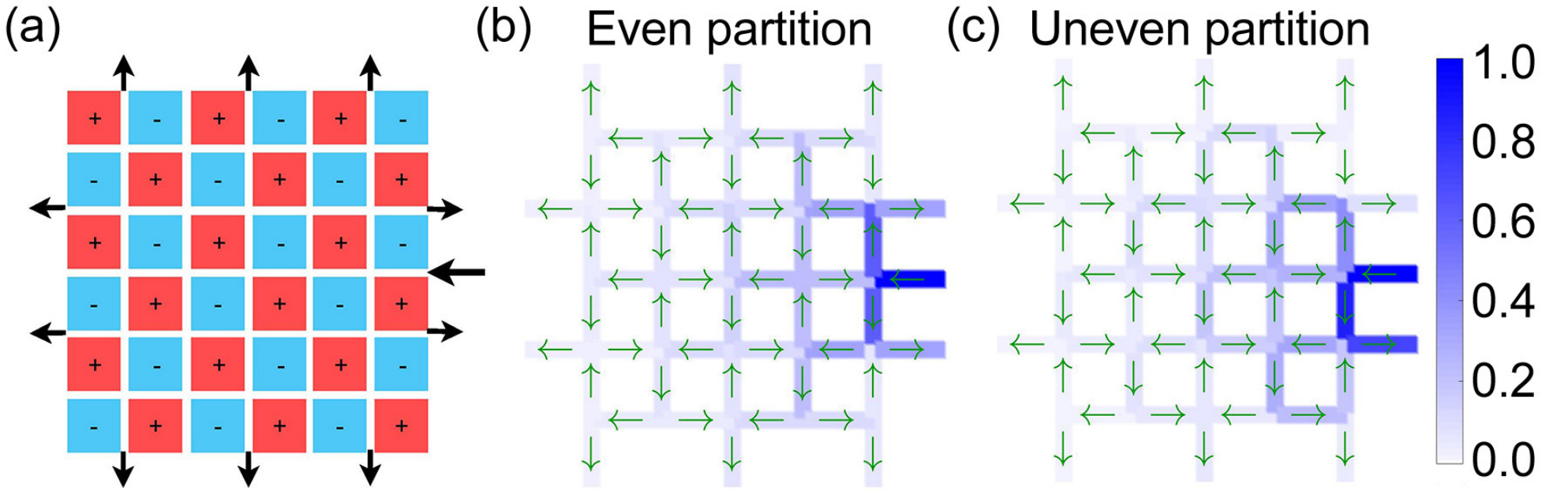}
  \caption{Partition and diffusion of currents in a regular network of zero-line nodes. (a) Schematic representation of the injected and outgoing currents and the opposite domains with different mass signs when the current is injected from a single right lead. The net current injection should be most efficient for same chirality currents contacted at alternating leads. (b) Current partition map and diffusion in a zero-line network when the current partitions evenly at each node. The alternating orbital moments at contiguous domains due to the loop currents approximately cancel each other. The current magnitude decreases substantially after a few partition steps. (c) Current partition and diffusion map when the partition is asymmetric. The asymmetry magnitudes of the loop currents at contiguous domains can give rise to a finite orbital magnetic moment.}
  \label{Fig4}
\end{figure}

\textbf{Current partition network}.
Networks consisting of multiple intersection of zero-line nodes are natural extensions of a single zero-line beam splitter setup whose collective behavior can dictate other collective current partition properties. The specific behavior of transport in a zero-line network will depend on system specific details such as the number of nodes, their distribution, and the way that the contact leads are connected to the zero lines. In the following, we have carried out an analysis of the current partition properties in a few specific setups of zero-line network that can be illustrative for understanding the behavior in a more generic case. As a first approximation, we considered in our discussions fixed current partition ratios expected from single zero-line beam splitter node calculations and neglecting quantum interference effects between the nodes, as well as tunneling to neighboring parallel zero lines that are potentially relevant at the nanoscale. The main observation is that the current injected into the network of zero lines will partition successively at the nodes and diffuse as the current magnitude decreases progressively.
This diffusion of the injected currents create loops in opposite directions at alternating mass domains as represented in Fig.~\ref{Fig4}. This observation suggests that enhancement of transport in a zero-line network can be achieved when the leads are connected at alternating zero-line leads such that the chirality of the current is the same, see Fig.~\ref{Fig4}. We expect that in this type of devices it should be possible to create alternating domains with opposite orbital moments by simply injecting and collecting a stream of current that passes through the network region, and a net orbital moment will be generated when current partition asymmetry is introduced by the introduction of a finite magnetic field and finite carrier density as discussed earlier. Analysis of transport when several zero-line leads are simultaneously contacted by the same electrode at each side of a rectangular device is discussed in the Supporting Information for the two different limiting cases of $2\times N$ and $M \times 2$ for long and wide networks, as well as $M \times M$ networks with $M,N=1$-$10$ as representative cases that illustrate the generic current transport and 
partition behavior in a zero-lines network.

\textbf{Summary}.
In summary, in this work we have studied the tunability of the current partition properties at a topological zero-line intersection node that can be designed by appropriately reversing the staggered sublattice potential in graphene, or more generically reversing the mass signs of a chiral two dimensional gas. We have identified new system parameters that can be used as control knobs for altering the current partition properties, which had not been discussed in earlier literature including the Fermi level, the ratio of the mass terms at each side of the zero line, and an uniform external magnetic field. In particular, the combination of Fermi level tuning in the presence of a finite perpendicular magnetic field allows to switch in opposite senses the current partition asymmetry by changing the Fermi level from electron-like to hole-like case. This possibility is particularly attractive as it provides a new way of routing the current by applying an appropriate electric gate potential in conventional field effect transistor setups. We have also studied the network of zero lines and shown that such network allows to efficiently diffuse the injected current by forming domains with alternating orbital moments of opposite signs, and maximum current injection happens when same chirality currents are injected at alternating leads. The total orbital moment can be obtained by summing the opposite contributions from alternating domains and they have maximum cancellation when we assume equal partition, while a finite magnetization should develop when there is an asymmetry in the partition. This behavior makes it possible to devise ultrathin nano-mesoscale electromagnet circuits where a finite magnetization can be induced by injecting current into a zero-line network with partition asymmetry, whose magnetization orientation should be switchable using external gates.

\newpage

\newpage

\textbf{Author Contributions}. Y.R. and K.W. contributed equally to this work.

\begin{acknowledgement}
We are grateful to Prof. A. H. MacDonald and Prof. Q. Niu for helpful discussions. This work was financially supported by the National Key R and D Program (2016YFA0301700), the China Government Youth 1000-Plan Talent Program, and NNSFC (Grant No. 11474265). S.A.Y. thanks the financial support by Singapore University of Technology and Design (SUTD-SRG-EPD2013062 and SUTD-T1-2015004). Work in Seoul has been supported by the Korean NRF under Grant Nos. NRF-2015K2A1B8066011 and NRF-2016R1A2B4010105. The Supercomputing Center of USTC is gratefully acknowledged for the high-performance computing assistance.
\end{acknowledgement}


\begin{thebibliography}{99}

\bibitem{QAHE1}
Haldane, F. D. M. \textit{Phys. Rev. Lett.} \textbf{1988}, 61, 2015.

\bibitem{TopologicalStates0}
Hasan, M. Z.; Kane, C. L. \textit{Rev. Mod. Phys.} \textbf{2010}, 82, 3045; and references therein.

\bibitem{TopologicalStates2}
Ren, Y. F.; Qiao, Z. H.; Niu, Q. \textit{Rep. Prog. Phys.} \textbf{2016}, 79, 066501; and references therein.

\bibitem{TopologicalStates1}
Weng, H. M.; Yu, R.; Hu, X.; Dai, X.; Fang, Z. \textit{Adv. Phys.} \textbf{2015}, 64, 227; and references therein.

\bibitem{TopologicalStates3}
Liu, C.-X.; Zhang, S.-C.; Qi, X.-L. \textit{Annu. Rev. Condens. Matter Phys.} \textbf{2016}, 7, 301; and references therein.

\bibitem{qiaoti}
Qiao, Z. H.; Tse, W. K.; Jiang, H.; Yao Y.; Niu Q. \textit{Phys. Rev. Lett.} \textbf{2011}, 107, 256801.

\bibitem{kanemele}
Kane, C. L.; Mele, E. J. \textit{Phys. Rev. Lett.} \textbf{2005}, 95, 226801.

\bibitem{QAHE2}
Liu, C. X.; Qi,  X. L.; Dai, X.; Fang, Z.; Zhang, S. C. \textit{Phys. Rev. Lett.} \textbf{2008}, 101, 146802.

\bibitem{QAHE3}
Yu, R.; Zhang, W.; Zhang, H. J.; Zhang, S. C.; Dai, X.; Fang, Z. \textit{Science} \textbf{2010}, 329, 61.

\bibitem{QAHE4}
Qiao, Z. H.; Yang, S. A.; Feng, W. X.; Tse, W.-K.; Ding, J.; Yao, Y. G.; Wang, J.; Niu, Q. \textit{Phys. Rev. B} \textbf{2010}, 82, 161414.

\bibitem{QAHE5}
Wang, Z. F.; Liu, Z.; Liu, F. \textit{Phys. Rev. Lett.} \textbf{2013}, 110, 196801.

\bibitem{QAHE6}
Garrity, K. F.; Vanderbilt, D. \textit{Phys. Rev. Lett.} \textbf{2013}, 110, 116802.

\bibitem{QAHE7}
Zhang, F.; Jung, J.; Fiete, G. A.; Niu, Q.; MacDonald, A. H. \textit{Phys. Rev. Lett.} \textbf{2011}, 106, 156801.

\bibitem{QAHEExp1}
Chang, C. Z.; Zhang, J.; Feng, X.; Shen, J.; Zhang, Z. \textit{et al.} \textit{Science} \textbf{2013}, 340, 167.

\bibitem{QAHEExp2}
Checkelsky, J. G.; Yoshimi, R.; Tsukazaki, A.; Takahashi, K. S.; Kozuka, Y.; Falson, J.; Kawasaki, M.; Tokura, Y. \textit{Nat. Phys.} \textbf{2014}, 10, 731.

\bibitem{QAHEExp3}
Kou, X.; Guo, S.-T.; Fan, Y.; Pan, L.; Lang M.; \textit{et al.}. \textit{Phys. Rev. Lett.} \textbf{2014}, 113, 137201.

\bibitem{QAHEExp4}
Chang, C.-Z.; Zhao, W.; Kim, D. Y.; Zhang, H.; Assaf, B. A.; Heiman, D.; Zhang, S.-C.; Liu, C. X.; Chan, M. H. W.; Moodera, J. S. \textit{Nat. Mat.} \textbf{2015}, 14, 473.

\bibitem{valleyhall}
Xiao D.; Yao W.; Niu Q. \textit{Phys. Rev. Lett.} \textbf{2007}, 99, 236809.
%


\bibitem{junghf}
 Jung J.; Zhang F.; MacDonald A. H. \textit{Phys. Rev. B} \textbf{2011}, 83, 115408.

\bibitem{Martin}
Martin I.;  Blanter Ya. M.; and  Morpurgo A. F. \textit{Phys. Rev. Lett.} \textbf{2008}, 100, 036804.

\bibitem{yao}
Yao W.; Yang S. A.; and Niu Q.  \textit{Phys. Rev. Lett.} \textbf{2009} 102, 096801.

\bibitem{semenoff}
Semenoff G.; W. Semenoff V. and Zhou F. \textit{Phys. Rev. Lett.} \textbf{2008}, 101, 087204.

\bibitem{Jeil}
Jung J.; Zhang F.; Qiao Z. H.; and MacDonald A. H. \textit{Phys. Rev. B} \textbf{2011}, 84, 075418.

\bibitem{eunah}
Vaezi A.; Liang Y.; Ngai D. H.; Yang L.; and Kim E.-A. A. \textit{Phys. Rev. X} \textbf{2013}, 3, 021018.

\bibitem{yongkuk}
Kim Y.; Choi K.; and Ihm J.; Jin H. \textit{Phys. Rev. B} \textbf{2014}, 89, 085429.

\bibitem{BiXintao}
Bi, X.; Jung, J.; Qiao Z. H. \textit{Phys. Rev. B} \textbf{2015} 92, 235421.

\bibitem{changhee}
Lee C. H.; Kim G.; Jung J.; and Min H.; \textit{Phys. Rev. B} \textbf{2016}, 94, 125438.

\bibitem{Highway}
Qiao, Z. H.; Jung, J.; Niu, Q.; MacDonald, A. H. \textit{Nano Lett.} \textbf{2011}, 11, 3453.

\bibitem{zhujun}
Li J.; Wang K.; McFaul K. J.; Zern Z., Ren Y.; Watanabe K.; Taniguchi T.; Qiao Z. H. and Zhu J. \textit{Nat. Nanotech.} \textbf{2016}, 11, 1060.

\bibitem{hujong}
Kim M.; Choi J.-H.; Lee S.-H.; Watanabe K.; Taniguchi T.; Jhi S.-H. and Lee H.-J. \textit{Nat. Phys.} \textbf{2016}, 12, 1022.

\bibitem{julong}
Ju L.; Shi Z.; Nair N.; Lv Y.; Jin C.; Velasco Jr. J.; O.-A. C.; Bechtel H. A.; Martin M. C.; Zettl A.; Analytis J. and Wang F. \textit{Nature} \textbf{2015} 520, 650.

\bibitem{partition}
Qiao, Z. H.; Jung, J.; Lin, C.; Ren, Y.; MacDonald, A. H.; Niu, Q. \textit{Phys. Rev. Lett.} \textbf{2014}, 112, 206601.

\bibitem{qiao-magneticfield}
Qiao Z. H.; Ren W; Wang J. and Guo H. \textit{Phys. Rev. Lett.} \textbf{2007}, 98, 196402.

\end{thebibliography}
\end{document}